\documentstyle{article}
\begin{document}
\hfuzz = 10pt
\newcommand{\Hol}{{\rm Hol}}
\newcommand{\tr}{{\rm tr}}
\newcommand{\Tr}{{\rm Tr}}
\newcommand{\Ad}{{\rm Ad}}
\newcommand{\ad}{{\rm ad}}
\newcommand{\Ascr}{{\cal A}}
\newcommand{\Dscr}{{\cal D}}
\newcommand{\Gscr}{{\cal G}}
\newcommand{\Fscr}{{\cal F}}
\newcommand{\Oscr}{{\cal O}}
\newcommand{\Lscr}{{\cal L}}
\newcommand{\Uscr}{{\cal U}}
\newcommand{\Tscr}{{\cal T}}
\newcommand{\Rscr}{{\cal R}}
\newcommand{\ba}{\begin{eqnarray}}
\newcommand{\ea}{\end{eqnarray}}
\newcommand{\lora}{\longrightarrow}
\newcommand{\reali}{{\hbox{{\rm I}\kern-.2em\hbox{\rm R}}}}
\newcommand{\complessi}{{\ \hbox{{\rm I}\kern-.6em\hbox{\bf C}}}}
\newcommand{\Hor}{{\rm Hor}}
\newcommand{\Aut}{{\rm Aut}}
\newcommand{\be}{\begin{equation}}
\newcommand{\ee}{\end{equation}}
\newcommand{\orbi}{{{\Ascr}/{\Gscr}}}
\newcommand{\sbi}{\mapstochar\joinrel\joinrel\times}

\begin{center}
{$BF$ Theories and 2-knots
\vskip 2em
Paolo Cotta-Ramusino \\
Dipartimento di Fisica, Universit\'a di Trento\\
and INFN, Sezione di Milano \\
Via Celoria 16, 20133 Milano, Italy\\
(email: cotta@milano.infn.it)}
\vskip 2em
{
Maurizio Martellini \\
Dipartimento di Fisica, Universit\'a di Milano\\
 and INFN, Sezione di Pavia \\
Via Celoria 16, 20133 Milano, Italy\\
(email: martellini@milano.infn.it)}

\end{center}

\begin{abstract}
We discuss the relations between
(topological) quantum field theories in 4 dimensions
and the theory of 2-knots (embedded 2-spheres in a 4-manifold).
The so-called $BF$ theories allow the construction of quantum
operators whose trace can be considered as the higher-dimensional
generalization of
Wilson lines for knots in 3-dimensions. First-order perturbative
calculations lead to higher dimensional linking numbers, and
it is possible to establish a heuristic relation between
$BF$ theories and Alexander invariants.
Functional integration-by-parts techniques allow the recovery
of an infinitesimal version of the Zamolodchikov tetrahedron equation,
in the form considered by Carter and Saito.
\end{abstract}

\section{Introduction}
In a seminal work,  Witten \cite{Witten1}   has shown that there is a very deep
connection between invariants of knots and links in a 3-manifold and the
quantum Chern--Simons field theory.

One of the questions that is possible to ask is whether there exists an
analog of  Witten's ideas in higher dimensions.  After all, knots and links
are defined in any dimension (as embeddings of $k$-spheres into a
$(k+2)-$dimensional space), and topological quantum field theories exist
in 4 dimensions \cite{Witten2,BBRT}.

The question, unfortunately, is not easy to answer.  On one hand, the theory
of higher dimensional knots and links is much less developed; the relevant
invariants are, for the time being,
more scarce then in the theory of ordinary knots and links.
As far as 4-dimensional topological field theories are concerned,
only the general BRST
structure has been thoroughly discussed; perturbative calculations
at least
in the non-abelian case, do not appear
to have been carried out. Even the quantum
observables have not been clearly defined (in the non-abelian case).
Moreover, Chern--Simons theory in 3 dimensions benefitted
greatly from the results of (2-dimensional)  conformal field theory.  No
analogous help is available for 4-dimensional topological
field theories.

In this paper we make some preliminary considerations and proposals
concerning 2-knots and topological field theories of the $BF$
type. Specifically we propose a set of quantum observables
associated
to surfaces embedded in 4-space. These observables can be seen as a
generalization of ordinary Wilson lines in 3 dimensions.  Moreover, the
framework in which these observables are defined fits with the
picture of 2-knots as ``movies" of ordinary links (or link-diagrams)
considered by Carter and Saito (see references below).
It is also possible to find heuristic arguments that may relate
the expectation values
of our observables in the
$BF$ theory with the Alexander invariants of 2-knots.

As far as perturbative calculations are concerned, they are much more
complicated then in the lower dimensional case.  Nevertheless
it is possible to recover a relation between first-order calculations
and (higher-dimensional) linking numbers, which parallels a similar
relation for ordinary links in a 3-dimensional space.

Finally, in 4 dimensions it is also possible to apply functional
integration-by-parts techniques.  In this case, these techniques produce
a solution of an infinitesimal Zamolodchikov tetrahedron
equation, consistent with the proposal made by Carter, Saito and
Lawrence.

\section{BF theories and their geometrical significance}

We start by considering a 4-dimensional Riemannian
manifold $M$, a compact Lie group $G$ and a principal bundle $P(M,G)$
over $M$.  If $A$ denotes a connection on such a bundle, then
its curvature will be denoted by $F_A$ or simply by $F$.
The space of all connections will be denoted by the symbol $\Ascr$
and the group of gauge transformations by the symbol $\Gscr$.
The group $G$, with Lie Algebra $Lie(G)$,  will be, in most cases,
the group $SU(n)$ or $SO(n)$.

Looking for possible topological actions
for a 4-dimensional manifold, we may consider
([2],[3]???) the (Gibbs measures of the):

\begin{enumerate}

\item ``Chern" action
\ba \exp \biggl(- \kappa \int_M \Tr_{\rho}(F\wedge F)\biggr), \label{Chern}\ea
and of the

\item  ``$BF$" action
\ba \exp \biggl(-\lambda \int_M \Tr_{\rho}(B\wedge F)\biggr). \label{BF} \ea

\end{enumerate}

In the formulas above $\kappa$ and $\lambda$ are coupling constants, $\rho$
is a given representation of the group $G$ and of its Lie algebra.
The curvature $F$ is a 2-form on $M$
with values in the associated bundle $P\times_{Ad} Lie(G)$
or, equivalently, a tensorial 2-form on $P$ with values in $Lie(G)$.

The field $B$ in eqn (\ref{BF}) is assumed (classically) to be a 2-form of the
same nature as $F$. We shall see though, that the
quantization procedure may force $B$ to take values in the
universal enveloping algebra of $Lie(G)$.
It is then convenient to discuss the geometrical aspects of
a more general $BF$ theory,
where $B$ is assumed to be a 2-form on
$M$ with values in the associated
bundle $P\times_{\Ad} \Uscr G$ where $\Uscr G$ is
the universal enveloping algebra
of $Lie G$. In other words $B$ is a section of the bundle:
$$\Lambda^2(T^*M)\otimes (P\times_{\Ad} \Uscr G).$$
The space of forms with values in $P\times_{\Ad} \Uscr G$
will be denoted by the symbol $\Omega^*(M,\Uscr \ad P)$:
this space naturally contains $\Omega^*(M,\ad P)$,
the space of forms with values in $P\times_{\Ad} Lie(G)$.

In order to obtain
an ordinary form on $M$ from an element of \break
$\Omega^*(M,\Uscr \ad P)$, one needs to take the trace with respect to a given
representation $\rho$ of $Lie(G)$.
Notice that the wedge product for forms
in $\Omega^*(M,\Uscr \ad P)$ is obtained by combining the product in
$\Uscr G$ with the exterior product for ordinary forms.

We now mention some examples of $B$-fields that have a nice
geometrical meaning:

\begin{enumerate}

\item  Let $LM\lora M$ be the frame bundle and let $\theta$ the
soldering form. It is a $\reali^n$-valued 1-form, given by the identity
map on the tangent space of $M$. For any given metric
$g$ we consider the reduced $SO(n)$-bundle of orthonormal frames $O_gM$. A {\it
vielbein} in physics is defined as the $\reali^n$-valued 1-form on $M$
given by
$\sigma^* \theta,$
for a given section $\sigma:M\lora O_gM.$
We now define the 2-form $B \in \Omega^2(M,\ad O_gM)$ as
$$B\equiv \theta \wedge \theta^T|_{O_gM},$$
where $^T$ denotes transposition.  In the corresponding
$BF$ action, the curvature $F$ is the curvature of a connection
in the orthonormal frame bundle $O_gM$. This action is known to be
(classically) equivalent to
the Einstein action, written in the vielbein formalism.

\item  The manifold $\Ascr$ is
an affine space with underlying vector space \break
$\Omega^1(M, \ad P)$.
We denote by $\eta$ the 1-form on $\Ascr$ given by the identity map on
the tangent space of $\Ascr$, i.e. on $\Omega^1(M, adP)$.
{}From this we can obtain the 2-form on $\Ascr$
\ba B\equiv \eta\wedge \eta , \label{B2} \ea
with values in $\Omega^2(M,\Uscr adP)$.
More explicitly, $B$ can be defined as
\ba B(a,b)(X,Y)|_{p,A}= a(X)b(Y)-a(Y)b(X)-b(X)a(Y)+b(Y)a(X),\label{B3} \ea
where $a,b\in T_A\Ascr;
\; X,Y\in T_pP. $

\end{enumerate}

We now want to show that the $BF$ action considered in the last
example is connected with the ABJ anomaly in 4 dimensions.
On the principal $G-$bundle
\ba P\times \Ascr\lora M\times \Ascr   \label{Gbundle} \ea
we can
consider the tautological connection defined by the following horizontal
distribution:
\ba   \Hor_{p,A}= \Hor_p^A \oplus T_A\Ascr    \label{tautconn} \ea
where $\Hor_p^A$ denotes the horizontal space at $p\in P$ with respect
to the connection $A$. The connection form of the tautological
connection is given simply by $A$, seen as a (1,0)-form
on $P\times \Ascr$. Its curvature form
is given by
\ba    \Fscr_{p,A}= (F_A)_p + \eta_A.     \label{Foftautconn} \ea
Let us now use $Q_l$ to denote an irreducible
ad-invariant polynomial on $Lie(G)$
with
$l$ entries.   Integrating  the Chern-Weil form corresponding
to such polynomials over $M$ we obtain (for $l=2,3,4$):
\ba \int_M Q_2 (\Fscr,\Fscr) =
k_2 \int_M Tr_{\rho} (F_A\wedge F_A) \label{Q2} \ea
\ba \int_M Q_3 (\Fscr,\Fscr,\Fscr) =
 k_3 \int_M Tr_{\rho} (B\wedge F_A) \label{Q3} \ea
\ba \int_M Q_4 (\Fscr,\Fscr,\Fscr,\Fscr) = k_4\int_M Tr_{\rho} (B\wedge
B) \label{Q4} \ea
where $B$ is defined as in eqns (\ref{B2},\ref{B3}),
and $k_l$ are normalization factors.
In eqn (\ref{Q4}) the wedge product also includes
the exterior product of forms on $\Ascr.$

As Atiyah and Singer \cite{AS}
pointed out, one can consider together
with \ref{Gbundle} the following
principal $G$-bundle
\footnote{Strictly speaking, in this case, one should consider only either the
space of irreducible connections and the gauge group, divided by its center, or
the space of all connections and the group of gauge transformations which
give the identity when restricted to a fixed point of $M$. We always
assume that one of the above choices is made.}:
\be   \displaystyle{{{P\times \Ascr}\over{\Gscr}}} \lora
M\times \displaystyle{{{\Ascr}\over{\Gscr}}}. \label{Gbundle2} \ee
Any connection on the bundle $\Ascr \lora
\orbi$ determines a connection on the bundle (\ref{Gbundle2}) and hence on
the bundle (\ref{Gbundle}). When we consider such
connection on (\ref{Gbundle}) instead of the tautological one, then the
left-hand side of
eqn (\ref{Q3}) becomes the term which generates, via the
so called ``descent equation",
the ABJ anomaly in 4 dimensions \cite{AS,BCRS}.
Hence the $BF$ action (\ref{Q3}) is a sort of gauge-fixed version of the
generating term for the ABJ anomaly.

In any $BF$-action one has to integrate forms defined on $M\times \Ascr$.
Hence we can consider different kinds of symmetries
\cite{AS,BS}:

\begin{enumerate}

\item
 ``connection-invariance": this refer to forms on
$M\times \Ascr$ which are closed: when they are integrated over
cycles of $M$, they give closed forms on $\Ascr$;
\item
 ``gauge invariance": this refer to forms which are defined over
$M\times \Ascr$, but are
pullbacks of forms defined over
$M\times \displaystyle{{{\Ascr}\over{\Gscr}}};$

\item
 ``diffeomorphism-invariance":
this can be considered whenever we have an action of
$Diff(M)$ over $P$ and over $\Ascr$.
In particular we can consider
diffeomorphism-invariance when $P$ is a trivial bundle,
or when
$P$ is the bundle of linear frames over $M$.
In the latter case
\be  \displaystyle{{{P\times \Ascr}\over{Diff(M)}}} \lora
\displaystyle{{{M\times \Ascr}\over{Diff(M)}}} \label{diffeoinv} \ee
is a principal bundle
\footnote{In fact
one should really consider only the group of diffeomorphisms of $M$
which strongly fix one point of $M$. Also, instead of $LM\times \Ascr$
it is possible to
consider
$\Oscr_M \sbi \Ascr^{metric}$, namely the space of all
orthonormal frames paired with the {\it corresponding } metric connections
(see e.g.\ \cite{BCRS}.};

\item diffeomorphism/gauge invariance: this refers to
the action of $\Aut P$, i.e.
the full group of automorphisms of the
principal bundle $P$ (including the automorphisms
which do {\it not} induce the
identity map on the base manifold). This kind of invariance
imply gauge invariance and diffeomorphism-invariance when the
latter one is defined.
\end{enumerate}

For instance the Chern action (\ref{Chern}) is both connection- and
diffeomorphism/ gauge-invariant, while the $BF$ action (\ref{BF}) is
gauge-invariant
and its connection-invariance depends on further specifications.
In particular the $BF$ action of example 2) is connection-invariant,
while the action of
example 1) is not.
More precisely, in example 1),
the integrand of the $BF$ action defines a closed 2-form
on $O_gM \times \Ascr$, i.e., we have
$$\delta_{\Ascr}\biggl(\int_M \Tr \{F \wedge [\theta
\wedge \theta^T]|_{O_gM}\}
\biggr)=0$$
only when the covariant derivative $d_A \theta$ is zero;
in other words, the critical connections are
the Levi-Civita
connections.

When $B$ is given by \ref{B2}, then
for any 2-cycle $\Sigma$ in $M$
$$\int_{\Sigma} \Tr_{\rho}B =\int_{\Sigma}Q_2( \Fscr\wedge\Fscr)$$
is a closed 2-form on $\Ascr$ and so we have a map
\be \mu: Z_2(M)\lora Z^2(\Ascr),     \label{mu} \ee
where $Z_2$ ($Z^2$) denotes 2-cycles (2-cocycles).
When we replace the curvature of the tautological
connection with the curvature of a connection in the bundle
(\ref{Gbundle2}), then we again obtain a map (\ref{mu}),
but now this map  descends to a map
\be   \mu: H_2(M) \lora
H^2(\displaystyle{{{\Ascr}\over{\Gscr}}}).\label{mu2} \ee
The map (\ref{mu2}) is the basis for the construction of the Donaldson
polynomials \cite{DK}.

As a final remark we would like to mention $BF$ theories in arbitrary
dimensions.  When $M$ is a manifold of dimension $d$, then
we can take the field $B$ to be
a $(d-2)$-form, again with values in
the bundle $P\times_{\Ad} \Uscr G$.
The form (\ref{B3}) makes then perfectly sense in any dimension.

A special situation occurs when the group $G$
is $SO(d)$. In this case there is a linear isomorphism
\be \Lambda^2(\reali^d) \lora Lie(SO(d)), \label{SO} \ee
defined by
$$e_i\wedge e_j \lora (E^i_j-E^j_i),$$
where $\{e_i\}$ is the canonical basis of $\reali^d$ and $E^i_j$ is the
matrix with entries $(E^i_j)^m_n=\delta_{m,i}\delta_{n,j}.$
The isomorphism (\ref{SO}) has the following properties:

\begin{enumerate}

\item
It is an isomorphism of inner product spaces, when the inner
product in $Lie(SO(d))$ is given by $(A,B)\equiv(-1/2) TrAB.$
\item The left action of $SO(d)$ on $\reali^n$ corresponds to the
adjoint action on $Lie(SO(d))$.

\end{enumerate}

\noindent
One can then consider a principal $SO(d)$-bundle and an associated bundle
$E$ with fiber $\reali^d$.
In this special case
we can consider a different kind of $BF$-theory,
where the field $B$ is assumed to be a $(d-2)$-form with values in the
$(d-2)$th exterior
power of $E$. In this case the corresponding $BF$ action is given by:
\be    \exp\bigl(-\lambda  \int_M B\wedge F\bigr),  \label{BFaction} \ee
where the wedge product combines the wedge product of forms on $M$ with the
wedge product in $\Lambda^*(\reali^d)$. The action (\ref{BFaction})
is invariant under gauge transformations.
When the $SO(d)$-bundle is the orthonormal bundle and the field $B$
is the $(d-2)$nd exterior power of the soldering form, then the
corresponding
$BF$ action (\ref{BFaction})
gives the (classical) action for gravity in $d$ dimensions, in the so-called
Palatini (first-order) formalism \cite{CJD,ARS}.

\section{2-knots and their quantum observables}
In Witten--Chern--Simons \cite{Witten1} theory we have a topological action on
a
3-dimensional manifold $M^3$, and the observables correspond to knots
(or links) in
$M^3$. More precisely to each knot we associate the trace of the
holonomy along the knot in a fixed representation of the group $G$,
or ``Wilson line.''
In 4 dimensions we have at our disposal the Lagrangians
considered at the beginning of the previous section.
It is natural to consider as observables, quantities (higher-dimensional
Wilson lines) related to 2-knots.

Let us recall here that while an
 ordinary knot is a 1-sphere embedded in $S^3$
(or $\reali^3$),
a 2-knot is a 2-sphere embedded in $S^4$ or
in the 4-space $\reali^4$. A generalized 2-knot in a four-dimensional
closed manifold $M$, can be defined as a {\it closed} surface $\Sigma$
embedded in $M$.
Two 2-knots (generalized or not) will be called equivalent if
they can be mapped into each other by a diffeomorphism connected to the
identity of the ambient manifold $M$.

The theory of 2-knots (and 2-links)
is less developed than the theory
of ordinary knots and links. For instance
it is not known whether one can have an
analogue of the Jones polynomial for 2-knots.  On the other hand,
one can define
Alexander invariants for 2-knots (see e.g.\ \cite{Rolfsen}).

The problem we would like to address here is whether there exists a
connection between 4-dimensional field theories and (invariants of)
2-knots. Namely, we would like to ask whether there exists a
generalization to 4 dimensions
of the connection established by Witten
between topological field theories and knot invariants in 3 dimensions.

Even though we are not able to show rigorously that a consistent set of
nontrivial
invariants for 2-knots can be constructed out of 4-dimensional field
theories, we can show that there exists a connection between $BF$ theories in
4 dimensions and 2-knots. This connection involves, in different places,
the Zamolodchikov tetrahedron equation as well as self-linking and
(higher-dimensional) linking
numbers.

In Witten's theory one has to functionally integrate an observable depending on
the given (ordinary)
knot (in the given 3-manifold) with respect to a
topological Lagrangian.
Here we want to do something similar, namely we want to
functionally integrate
different observables depending on a given embedded surface $\Sigma$ in
$M$ with respect to a path-integral measure given by the $BF$-action.

The first (classical) observable we want to consider is
\be  \Tr_{\rho}\exp\bigl(\int_{\Sigma} \Hol(A,\gamma_{y,x}) B(y)
\Hol(A,\gamma_{z,y})\bigr),\quad y\in \Sigma. \label{observable1} \ee
Here by $\Hol(A,\gamma_{y,x})$ we mean the holonomy with respect to the
connection $A$ along the path $\gamma$ joining two given
points $y$ and $x$ belonging to $\Sigma$\footnote
{We assume to have chosen once for all a reference section for the
(trivial) bundle $P|_{\Sigma}$. Hence the holonomy along a path
is defined by the comparison of the horizontally
lifted
path and the path lifted through the reference section.}.
Expression (\ref{observable1}):

\begin{enumerate}
\item  depends on the choice of a map assigning to each
$y\in \Sigma$ a path $\gamma$ joining $x$ and $z$ and passing through $y$;
more precisely, it depends on the holonomies along such paths.
One expects that the functional integral over the space of all connections
will integrate out this dependency. More precisely, we recall that the
(functional) measure over the space of connections is given, up to a Jacobian
determinant, by the measure over the paths over which the holonomy is
computed. See in this regard the analysis of the Wess-Zumino-Witten
model made by Polyakov and Wiegmann \cite{PW}.

\item
is gauge invariant if we consider only gauge transformations
that are the identity at the points $x$ and $z$. In the special case when
$z$ and $x$ coincide,
then \ref{observable1} is gauge invariant without any restriction.
\end{enumerate}

The functional integral of (\ref{observable1})
can be seen as the vacuum expectation value of the trace (in the
representation $\rho$) of
the {\it quantum surface operator} denoted by $O(\Sigma; x,z)$.
In other words we set:
\be O(\Sigma;x,z)\equiv \exp\bigl(\int_{\Sigma} \Hol(A,\gamma_{y,x}) B(y)
\Hol(A,\gamma_{z,y})\bigr),\quad y\in \Sigma;  \label{observable2} \ee
where, in order to avoid a cumbersome notation, we did not write the explicit
dependency on the paths $\gamma$, and
we did not
use different symbols for
the holonomy and its quantum counterpart ({\it quantum holonomy}).
Moreover in (\ref{observable2})
a time-ordering symbol has to be understood.

We point out that we will also allow
the operator (\ref{observable2}) to be defined for any embedded
surface $\Sigma$ {\it closed or with boundary}. In the latter case
the points $x$ and $z$ are always supposed to belong to the boundary.

As far as the role of the paths $\gamma$
considered above is concerned, we recall that, in a 3-dimensional theory with
knots, a
framing is a (smooth) assignment of a tangent vector to each point of
the knot. In four dimensions, instead, we assign a curve to each point of the
2-knot. We will refer to this assignment as a {\it (higher-dimensional)
framing}.  We will also speak of a {\it framed 2-knot}.

\section{Gauss constraints}
In order to study the quantum theory corresponding to (\ref{BF}), we first
consider the canonical quantization approach.
We take the group $G$ to be $SU(n)$ and consider its fundamental
representation, hence we write the field $B$ as
$\hat B + \sum_a B^a R^a$. Here $\hat B$ is a
multiple of $1$ and $\{R^a\}$  is an orthonormal basis of $Lie(G)$.
In the $BF$ action we can therefore disregard
the $\hat B$ component of the field $B$.

We choose a time-direction $t$
and write in local coordinates $\{t,x\}\equiv \{t,x^1,x^2,x^3\}$:
$$d= d_t + d_x,\quad
A= A_0 dt +A_x= A_0 dt + \sum A_i dx^i\quad (i=1,2,3), $$
$$F_A =  \sum_i (d_tA_i)\wedge dx^i +(d_xA_0)\wedge dt+
\sum_i [A_i,A_0]dx^i\wedge dt +  \sum_{i<j} F_{i,j}dx^i
\wedge dx^j$$
$$B=\sum_{i<j} B_{i,j}dx^i\wedge dx^j + B_{i,0} dt\wedge dx^i.$$
In local coordinates the action is given by:
\ba \Lscr &\equiv& \int_M Tr(B\wedge F_A) \nonumber \\
&\approx& \int_{M^3 \times I}
\sum_{i,j,k}Tr (B_{i,j} d_tA_k +
B_iF_{j,k}dt+ A_0(DB)_{i,j,k}) dx^i\wedge dx^j\wedge dx^k,
\label{BFaction2} \ea
where $D$ denotes the covariant 3-dimensional derivative with respect
to the connection $A_x$.

We first consider the {\it pure} $BF$-theory, namely, a $BF$ theory
where no embedded surface is considered.
We perform a Legendre transformation for the Lagrangian
$\Lscr$;
the conjugate momenta to the fields $B_i$ and $A_0$,
namely $\displaystyle{{{\partial \Lscr}\over{\partial (\partial_tB_i)}
}}$ and
$\displaystyle{{{\partial \Lscr}\over{\partial (\partial_tA_0)}
}}$, are zero, so we have the primary
constraints:
\be \displaystyle{{{\partial \Lscr}\over{\partial B_{i,0}} }}=\sum_{j,k}
\epsilon_{i,j,k}
(F_{A_x})_{j,k}=0, \quad
\displaystyle{{{\partial \Lscr}\over{\partial A_0}
}}= \sum_{i,j,k} \epsilon_{i,j,k} (DB)_{i,j,k}=0. \label{constraints} \ee
We do not have secondary constraints since the
Hamiltonian is zero, hence the time
evolution is trivial.
At the formal quantum level the constraints
(\ref{constraints}) will be written as:
\be (F_{A_x})_{op}|\phi_{\hbox {physical}}\rangle =0,\quad
(DB)_{op}|\phi_{\hbox {physical}}\rangle =0,\label{qconstraints} \ee
where $_{op}$ stands for ``operator" and the vectors
$|\phi_{\hbox {physical}}\rangle$ span the physical Hilbert space of the
theory.

We now consider the expectation value of
$\Tr_{\rho} O(\Sigma;x,x)$ (see definition (\ref{observable2})).
We still want to work in the canonical formalism,
the result will be that instead of the constraints
(\ref{constraints},\ref{qconstraints}),
we will have {\it a source} represented by the assigned surface $\Sigma$.

In order to represent the surface operator (\ref{observable2})
as a source-action term to be added to the $BF$-action,
we assume that we have a current (singular 2-form) $J_{\rm sing}$
concentrated on the surface $\Sigma$, of the form:
\[J_{\rm sing} = \sum_a J^a_{\rm sing} R^a,\]
where again $R^a$ is an orthonormal basis of $Lie(G).$
We assume in general that $J_{\rm sing}$ can take
values in
$\Uscr G$
\footnote{Hence by definition
$J_{\rm sing}$ has no component in $\complessi \subset
\Uscr(G)$.}.

We are now in position to write the operator
(\ref{observable2}) as:
\be O(\Sigma; z,z) = \exp \biggl(\int_M \tilde{\tr}[\Hol(A,\gamma_{y,z}) B(y)
\Hol(A,\gamma_{z,y}) \wedge J_{sing}] \biggr),
\label{observable3} \ee
where for any $A,A' \in \Uscr G, \; A=\sum_a A^a R^a,\; A'=\sum_a {A'}^a  R^a$
we have set \footnote{In particular
$\tilde{tr}$ coincides with $tr$ when $A,A' \in Lie(G).$}.
$$\tilde{tr}(AA') = \sum_a A^a{A'}^a$$
{}From (\ref{observable3}) it follows that the component of $J_{sing}$ in
$Lie(G)$ does
not give any significant contribution. We now
assume $J_{sing}\in
Lie(G) \otimes Lie(G)$. In this way we neclect
possible higher order terms in $Lie(G)^{\otimes k}
\subset \Uscr G $, but this is consistent with
our semi-classical treatment.

With the above notation and taking into account the $BF$ action as
well, the observable to be functionally integrated
is:
\be \Tr_{\rho}\;\exp \biggl(\int_M \tilde{tr}\bigl\{[\Hol(A,\gamma_{y,z}) B(y)
\Hol(A,\gamma_{z,y})] $$ $$\wedge [J_{sing} -
\Hol(A,\gamma_{z,y})^{-1} F(y)
\Hol(A,\gamma_{y,z})^{-1}] \bigr\}\biggr).\label{funcint} \ee
At this point the introduction of the singular current allows
us to formally represent a theory with sources concentrated on
surfaces, as a pure $BF$ theory with a new ``curvature" given by the difference
of the previous curvature and the currents associated to such sources.
{}From eqn (\ref{observable3}) we conclude moreover that $J_{\rm sing}$ as a
2-form
should be such that $J_{\rm sing}(y) \wedge d^2\sigma(y) \neq 0,\; y\in
\Sigma$,
where $d^2\sigma(y)$ is the surface  2-form of $\Sigma$.

Now we choose a time direction and
proceed to an
analysis of the Gauss constraints as in the pure $BF$
theory.  We assume that locally the four manifold $M$ is given
by $M^3 \times I$ ($I$ being a time-interval)
and the intersection of the 3-dimensional
manifold $M^3 \times \{t\}$ with the surface $\Sigma$ is an ordinary
link $L_t$.

In a {\it neighborhood of $y\in \Sigma$} the current $J_{\rm sing}$ will
look like:
\be J^a_{\rm sing}(x;y) =\delta_{\Pi}^{(2)}(x-y) R^a \;dx^1_{\Pi}\wedge
dx^2_{\Pi},
\quad y \in
\Sigma, \label{Jsing} \ee
where $\Pi$ is a plane orthogonal to $\Sigma$ at $y$ (with coordinates
$x^1_{\Pi}$ and $x^2_{\Pi}$ ) and $\delta^{(n)}$ denotes the $n$-dimensional
delta function.

The Gauss constraints imply that in the above neighborhood of $y\in \Sigma$
we have:
\be F^a(x) = \Hol(A,\gamma_{z,y}) J_{sing}^a(x;y)
\Hol(A,\gamma_{y,z});\; y\in \Sigma, x\in \Pi
\label{F} \ee
where $\gamma$ is a loop in $\Sigma$
based at $z$ and passing through $y$.
The right-hand side of (\ref{F}) is also equal to
$$ \displaystyle{{{\partial
\Hol(A,\gamma_{z,y,z})}\over{\partial
A^a(x)}}}
;\; y\in\Sigma, x\in \Pi. $$

Notice that the previous analysis implies that the
components of the curvature
$F^a$ (and consequently the components
of the connection $A^a$) must take values
in a noncommutative algebra, at least when
sources are present.
More precisely the
components $F^a$ (as well as $A^a$) should be proportional to
$R^a \in Lie(G)$.
Given the fact that $B^a$ and $A^a$
are conjugate fields, we can conclude that the (commutation
relations of the) quantum theory will
require $B^a$ to be proportional to $R^a$ as well.

As in the lower-dimensional case
(ordinary knots (or links) as sources in a 3-dimensional theory),
\cite{GMM})
the restriction of the connection $A$ to the plane $\Pi$
looks (in the approximation for which
$\Hol(A,\gamma_{z,x,z}) =1$)
like
\be   A^a(x)|_{\Pi}= R^a \;d \;\log(x-y)     \label{A} \ee
for a given $y\in \Sigma$ (here $d$ is the exterior derivative).

In order to have some geometrical insight into the above connection,
we consider a linear approximation,
where the surface $\Sigma$ can be approximated
by a collection
of 2-dimensional planes in $\reali^4$, i.e., by a collection of
hyperplanes in $\complessi^2$.
These hyperplanes, and the hyperplane $\Pi$ considered above,
are assumed to be in general position. This means that the intersection of
the hyperplane $\Pi$ with any other hyperplane is given by a point.

The quantum connection (\ref{A}) on $\Pi$ gives a representation of the
first homotopy group of the manifold of the
arrangements of hyperplanes (points) in $\Pi$, i.e., more simply, of the
(pure) braid group \cite{Witten1,Kohno}.
It is then natural to ask whether the same is true in 4 dimensions, namely
whether the 4-dimensional connection, corresponding to the
(critical) curvature (\ref{F}) (in the linear
approximation defined above), is related to the higher braid group
\cite{MS}.
We will discuss this point further elsewhere;
let us
mention only that the 4-dimensional critical connection is related to
the existence of a higher dimensional
version of the Knizhnik-Zamolodchikov equation
associated with the representation of the higher braid groups,
as suggested by Kohno \cite{Kohno2}.

\section{Path integrals and the Alexander invariant of a 2-knot}
In the previous section we worked specifically in the
canonical approach. When instead we consider a covariant approach,
and compute the expectation values, with respect to the $BF$ functional
measure only,
we can write \footnote{We omit the ghosts and gauge fixing terms
\cite{BBRT}, since
they are not relevant for a rough calculation of (\ref{expect1}.}:
\[  \langle \Tr_{\rho} \;O(\Sigma;x,x)\rangle = \]
\be \int \Dscr A \Dscr B \;\Tr_{\rho}
\; \exp\biggl(\tilde{\tr}\int_M
\bigl[B(x) \wedge (F(x) -\Hol(A,\gamma_{x,z}) J_{\rm sing}(x)
\Hol(A,\gamma_{z,x}))\bigr]\biggr)\label{expect1} \ee
Again, in the approximation in which $\Hol(A,\gamma_{x,z})=1,$
we can set
\be \langle \Tr_{\rho}\;O(\Sigma;x,x) \rangle
\approx \int \Dscr A \Dscr B \;\Tr_{\rho}
\; exp\biggl(\tilde{\tr} \int_M
\bigl[B(x) \wedge (F(x) - J_{\rm sing}(x))\bigr]\biggr). \label{expect2} \ee
A rough (and formal) estimate of the previous expectation value (\ref{expect2})
can be given as follows.

By integration over the $B$ field we obtain
something like
\be
\langle \Tr_{\rho}\;O(\Sigma;x,x) \rangle
\approx \dim(\rho)\int \Dscr(A) \delta (F -J_{\rm sing}), \label{expect3} \ee
using a functional Dirac delta.
By applying the standard formula for Dirac deltas evaluated on
composite functions, we heuristically obtain :
\be \langle \Tr_{\rho}O(\Sigma;x,x) \rangle \approx
\dim(\rho)|\det(D_{A_0}|_{M\backslash \Sigma})|^{-1}, \label{expect4} \ee
where
$\det(D_{A_0}|_{M\backslash \Sigma})$
denotes the (regularized) determinant
of the covariant derivative operator with respect to a background {\it flat}
connection
$A_0$ on the space $M\backslash \Sigma$.

It is then natural to interpret the expectation value (\ref{expect4}) (with
the Faddeev-Popov ghosts included)
as the analytic (Ray-Singer) torsion for the complement of
the 2-knot $\Sigma$ (see \cite{S,BT}).
This torsion is related to the Alexander invariant of the 2-knot
\cite{Turaev}.

\section{First-order perturbative calculations and \newline
 higher-dimensional linking numbers}

Let us now consider the expectation value
of the quantum surface observable
$\Tr_{\rho}\; O(\Sigma;x,x),$ {\it with respect to the
total $BF$ $+$ Chern
action}. The two fields involved in the quantum surface operator
are $A^a_{\mu}$ (the connection) and $B^a_{\mu,\nu}$ (the $B$-field)
with $a=1,\cdots, \dim(G)$ and
$\mu,\nu = 1,\cdots,4.$
The connection $A$ is present in the quantum surface operator via the
holonomy of the paths $\gamma_{y,x}$ and $\gamma_{x,y}$ ($y\in \Sigma$)
as in definitions (\ref{observable1}) and (\ref{observable2}).

At the first-order approximation in perturbation theory (with a background
field and covariant gauge),
we can write the holonomy as
\[ \Hol(A,\gamma_{y,x})\approx 1 + \kappa
\int_{\gamma_{y,x}} \sum_{\mu}A_{\mu}(z)dz^{\mu} +\cdots. \]
At the same order,
the only relevant part of the $BF$ action is the kinetic part
($B\wedge dA$),  so we
get the following  Feynman propagator:
\be
\langle A^a_{\mu}(x) B^b_{\nu\rho}(y)\rangle
=\displaystyle{{{2i}\over{\lambda}}
\delta^{ab}{1\over{4\pi^2}}\sum_{\tau} \epsilon_{\mu\nu\rho\tau}
{{(x^{\tau}-y^{\tau})}\over {|x-y|^4}}}.\label{propagator} \ee
In order to avoid singularities, we perform the usual point-splitting
regularization. This is tantamount as lifting the loops $\gamma_{x,y,x}$
in a neighborhood of $y$. The lifting is done in a direction
which is normal to the surface $\Sigma$.

The complication here is that the loop $\gamma$ itself
(based at $x$ and passing
through $y$) depends on the point $y$.
While the general case seems difficult to handle, we can
make simplifying
choices
which appear to be legitimate from the point of view
of the general framework of quantum field theory.
For instance we can easily assume that when we assign to a point $y$ the loop
$\gamma_{x,y,x}$ based at $x$ and passing through $y$,
then the same loop will be assigned to any other
point $y'$ belonging to $\gamma$.

Moreover we can consider an  arbitrarily
fine triangulation $T$ of $\Sigma$  and take into consideration
only one loop $\gamma_T$ which has non empty
intersection with each triangle of $T$.
This approximation will break gauge-invariance, which will be,
in principle,
recovered only in the limit when the size of the triangles go to zero.
(For a related approach see \cite{Arefeva}.)
The advantage of this particular approximation is that we have
only to deal with one loop $\gamma_T$, which by point-splitting
regularization is then completely lifted from the surface.  We call this
lifted path $\gamma_T^r$, where $r$ stands for regularized.

Finally we get (to first-order approximation in perturbation theory
and up to the normalization factor $\langle 1 \rangle$
\footnote{While the normalization factor for a pure (4-dimensional)
$BF$ theory appears
to be trivial \cite{BT}, in a theory with a combined $BF$ $+$ Chern
action we have a contribution coming from the Chern action. This contribution
has
been related to the first Donaldson invariant of $M$ \cite{Witten2}.
}):
\[     \langle Tr_{\rho}\;O(\Sigma;x,x)\rangle \approx \]
\be
\dim(\rho) \bigg[1+ C_2(\rho) \displaystyle{{{2i\kappa}\over{\lambda}}
{1\over{4\pi^2}}}\sum_{\mu,\nu,\rho,\tau}
\epsilon_{\mu\nu\rho\tau}
\int_{\Sigma} dx^{\mu} dx^{\nu}
\int_{\gamma_T^r}dy^{\rho} \displaystyle
{{{(x^{\tau}-y^{\tau})}\over {|x-y|^4}}}+ \cdots )\biggr],
\label{Cherncontrib} \ee
where $C_2(\rho)$ is the trace of the quadratic Casimir operator in the
representation $\rho.$
When $\Sigma$ is a 2-sphere, then (\ref{Cherncontrib}) is given by:
\[ \dim(\rho) \biggr[1+ C_2(\rho) \displaystyle{{{2i\kappa}\over{\lambda}}}
\Lscr(\gamma^r_T, \Sigma) +\cdots \biggr] \]
where $\Lscr(\gamma^r_T, \Sigma)$ is the (higher-dimensional)
{\it
linking number} of
$\Sigma$ with $\gamma^r_T$ \cite{HS}.

\section{Wilson ``channels"}
Let $\Sigma$ again represent our 2-knot (surface embedded in the 4-manifold
$M$).
Carter and Saito \cite{CS,CS2} inspired by a
previous work by Roseman \cite{Roseman} describe a
2-knot $\Sigma$ (in fact an embedded 2-sphere)
in 4-space by
the 3-dimensional analogue of a knot-diagram: they project down $\Sigma$
to a 3-dimensional space, keeping track of the over- and under-crossings.
One of the coordinates of this 3-space is interpreted as ``time."
The intersection of this 3-dimensional
diagram of a 2-knot with a plane (at a fixed time)
gives an ordinary link-diagram,  while the collection of all these link
diagrams at different times (``stills") gives a ``movie" representing
the 2-knot.

In the Feynman formulation of field theory, we start by considering
the surface
$\Sigma$ embedded in the 4-manifold $M$ directly, not its projection.
At the local level $M$ will be given
by $M^3 \times I$ ($I$ being a time-interval), so that
the intersection of the 3-dimensional
manifold $M^3 \times \{t\}$ with $\Sigma$ gives an ordinary
link $L_t$ (for all times $t$) in $M^3$.
In order to connect quantum field theory
with the standard theory of 2-knots,
we assume
more simply that $M$ is the 4-space $\reali^4$ and that
$\Sigma$ is a 2-sphere. One coordinate
axis in $\reali^4$ is interpreted as time, and the intersection of
$\Sigma$ with 3-space (at fixed times) is given by ordinary links,
with the possible exception of some critical values of the time parameter.

In other words, we have a space-projection
$\pi_t: \reali^4\lora \reali^3 $ for each time
$t$, so that
\[L_t(\Sigma)\equiv \pi^{-1}_t(\reali^3)
\cap \Sigma\]
 is, for non-critical times, an ordinary link.
For any time interval $(t_1,t_2)$ we will use also the notation
$\Sigma_{(t_1,t_2)}
\equiv \bigcup_{t\in(t_1,t_2)} \{\pi^{-1}_t(\reali^3)\cap \Sigma\}$.

We consider a set $\Tscr$
of (noncritical) times $\{t_i\}_{i=1,\cdots,n}.$ We denote the components
of the
corresponding links $L_{t_i}$ by the symbols
$K_i^{j(i)}, \;j=1,\cdots,s(i)$.

Here and below we will write $j(i)$ simply to denote the fact that
he index $j$ ranges over a set depending on $i$.
We choose one base-point $x^{j(i)}_i$ in each knot $K^{j(i)}_i$ and we denote
by $\Sigma^{j(i),j(i+1)}_i$ the surface contained in $\Sigma$
whose boundary includes $K^{j(i)}_i$ and $K^{j(i+1)}_{i+1}.$
We then assign a framing to $\Sigma^{j(i),j(i+1)}_i$ as follows:
for each point $p$ in the interior of $\Sigma^{j(i),j(i+1)}_i$ we associate
a path with endpoints $x^{j(i)}_i$ and $x^{j(i+1)}_{i+1}$ passing through
$p$.
To the surface $\Sigma^{j(i),j(i+1)}_i$ {\it with boundary components}
$K^{j(i)}_i$ and $K^{j(i+1)}_{i+1},$
we associate the operator:
\be \Hol(A, K^{j(i)}_i)_{x{^{j(i)}_i}}\;
O(\Sigma{^{j(i),j(i+1)}_i}; x{^{j(i)}_i}, x{^{j(i+1)}_{i+1}})\;
\Hol(A,K{^{j(i+1)}_{i+1}})_{x{^{j(i+1)}_{i+1}}},\label{surfaceop} \ee
where the subscript in the symbol $Hol(A,\ldots)$ denotes the base-point.

We refer to the 2-knot $\Sigma$
equipped with the set $\Tscr$ of times as the {\it (temporally) sliced
2-knot.}
For the sliced 2-knot, we define
the {\it Wilson channels} to be
the operators of the form

\[  O(\Sigma_{(-\infty,t_1)})
\;\Hol(A, K^1_1)_{x^1_1}\;
\cdots \Hol(A,K^{j(i)}_i)_{x^{j(i)}_i} \]
\be
O(\Sigma{^{j(i),j(i+1)}_i}; x{^{j(i)}_i}, x{^{j(i+1)}_{i+1}})\;
\Hol(A,K^{j(i+1)}_{i+1})_{x^{j(i+1)}_{i+1}}\cdots O(\Sigma_{(t_n,+\infty)};
x^{j(n)}_n,+\infty).\label{Wilsonchannel} \ee

We assume the following requirements for the Wilson channels:

\begin{enumerate}

\item
the operators which appear in a Wilson channel are chosen in such a
way that a surface operator is sandwiched between knot-operators only if
the relevant knots belong to the boundary of the surface;

\item the paths that are included
in the surface operators
$O(\Sigma{^{j(i),j(i+1)}_i})$
are not allowed to touch the boundaries of the surface,
except at the initial and final point.
\end{enumerate}

{\it Finally, to the ``sliced" 2-knot we associate the product of the
traces of  all possible Wilson channels. } This product can be seen as a
possible higher-dimensional generalization of the Wilson operator
for ordinary links in 3-space that was considered by Witten.

We would like now
to compare the Wilson channel
operators associated to a sliced 2-knot $\Sigma$ with the
operator
\be     O(\Sigma;-\infty, +\infty)    \label{operator}   \ee
considered in Section 5
(no temporal slicing involved).
The main difference concerns the prescriptions that are given
involving the paths $\gamma$'s that enter the definition of (\ref{operator})
(or, equivalently,
the framing of the 2-knot).
The Wilson channels are obtained from
(\ref{operator}) by:

\begin{enumerate}
\item choosing a finite set of times
$\Tscr \equiv \{t_i\}$, and

\item requiring
the paths $\gamma$'s
{\it to follow one of the components of the link $L_t$, for each
$t\in \Tscr$.}   \end{enumerate}

Finally, recall that in the Wilson channel operators,
each of the paths $\gamma$'s is forbidden to
follow two separate components of the same link
$L_t$, for any time $t\in \Tscr.$
This last condition can be understood by noticing that in order to
follow both the $j$th and the $j'$th
components of the link $L_{t_i}$,
for some $i, j(i), j'(i)$, one path
should follow the $j$th component, then
enter the surface
$\Sigma{^{j(i),j(i+1)}_i}$ (which is assumed
to be equal to $\Sigma{^{j'(i),j''(i+1)}_i}),$
and finally go {\it back} to the $j'$th component, thus contradicting  a
(local) causality condition.

\section{Integration by parts}
The results of the previous section imply that,
given a temporally sliced 2-knot
$\Sigma$, we can construct quantum operators by considering all the
Wilson channels relevant to surfaces $\Sigma{^{j(i),j(i+1)}_i},\; i\leq l
\;(i\geq l)$
and links $L_{t_i}$ for $i< l \;(i>l).$ Let us denote these
quantum operators by the symbols
\be W^{\rm in}(\Sigma, t_l),\quad (W^{\rm out}(\Sigma, t_l)).\label{W} \ee
They can be seen
as a sort of convolution, depending on $\Sigma$,
of all the quantum holonomies associated
to the links $L_{t_i}$ for $i< l \;(i>l).$

Ordinary links are closure of braids, and one may
describe the surface which generates the different
links at times $\{t_i\},$ as a sequence (movie) of braids \cite{CS3}.
But a link is also obtained by closing a tangle, and the
time-evolution of a link (represented by a surface)
can also be described in terms of 2-tangles, i.e. movies
of (ordinary) tangles \cite{Fischer}.
2-tangles form a (rigid, braided) 2-category \cite{Fischer}.
The 2-tangle approach (with its 2-categorical
content) may reveal itself as a useful one
for quantum field theory.

The initial tangle is called the source and the final tangle is called
the target.
In our case, say, the source represents
$L_{t_l}$ while the target represents $L_{t_{l'}},$ for $l'>l$.
The quantum counterpart of the the source and target is represented
by the quantum holonomies associated to the corresponding links,
while the quantum counterpart
of the 2-tangle is represented by the
quantum surface operator relevant to the surface
$\Sigma_{(t_l,t_{l'})}$.

One can think to the quantum surface operators as acting on
quantum holonomies (by convolution), but
it is difficult to do {\it finite} calculations and write
this action explicitly. What we can do instead is
to make some
{\it infinitesimal} calculations using integration-by-parts techniques
in Feynman path integrals.

Generally speaking, in a finite time evolution $t_l\lora t_{l'}$, the
quantum surface operator (defined by the surface with boundaries
$L_{t_l}$ and $L_{t_{l'}}$) will map the operator $W^{\rm in}(\Sigma, t_l)$
into the operator $W^{\rm in}(\Sigma, t_{l'}).$
Let us now consider what can be seen as the ``derivative" of this map.
Namely, let us see what happen when we consider the quantum
operator corresponding to a
surface bordering two links $L_{t_l}$ and $L_{t_l}^{\prime}$ that differ by
an elementary change.
{\it What the functional integration-by-parts rules will show, is that one can
interchange a variation of the surface operator with a variation of the link
(and consequently of its quantum holonomy).} So in order to compute the
``derivative" of the surface operator, we may consider
variations of the link
$L_{t_l}$.

In order to describe
elementary changes of links, recall that
braids or tangles (representing links at a fixed time)
can be seen as collection of oriented strings, while 2-tangles describe the
interaction of those strings \cite{Zam}.
We consider now two links that differ from each other
only in a (small) region involving three
strings. Let us number these strings with numbers $1,2,3$. The string
$1$ crosses under the string $2$, then the string $2$ crosses under the
string 3 and the string 1 crosses under the string 3. We denote by the symbols
$a$ and $b$ the part of each string that precedes and, respectively, follows
the first crossing, as shown in Fig.\ 1.

\begin{figure}[tb]
\begin{center}
\setlength{\unitlength}{1cm}
\begin{picture}(4,4)
\end{picture}
\end{center}
\caption{Three strings}
\end{figure}

The integration-by-parts rules for the $BF$ theory are as follows.
We consider a knot $K_l^{j(l)}$, which we
denote here by $C$ in order to simplify the notation,
and a point $x\in C$.
By the symbol
$\delta_x$ we mean the variation (of $C$)
obtained by inserting an infinitesimally
small loop
$c_x$ in $C$ at the point $x$.  We obtain:
\[   \delta_x \;\bigl(\Hol(A,C)_x\bigr)
\exp \bigl(-\lambda \int_M \Tr(B\wedge F)\bigr) \]
\[  =  \sum_{\mu\nu a}
d^2c_x^{\mu\nu} \Hol(A,C)_x \;F^a_{\mu\nu}\;R^a\;
exp \bigl(-\lambda \int_M \Tr(B\wedge F)\bigr)
\]
\be
=\displaystyle{{-1\over{2\lambda}}}\sum_{\mu\nu\rho\tau a}
\epsilon_{\mu\nu\rho\tau}\;d^2 c_x^{\mu\nu}
\;\exp \bigl(-\lambda \int_M \Tr(B\wedge F)\bigr)
\Hol(A,C)_x \; R^a \;\displaystyle{{{\delta}\over{\delta B^a_{\rho\tau}(x)}}},
\label{intbyparts} \ee
where  $d^2c_x$ denotes the surface 2-form of the surface
bounded by the infinitesimal loop
$c_x$, $R^a$ denotes as usual an orthonormal basis of $Lie(SU(n))$
and the trace is meant to be taken in the fundamental representation.
The second
equality has been obtained by {\it functional integration by parts};
in this way we
produced the functional derivative with respect to the $B$-field.
The integration-by-parts rules of the $BF$ theory in 4 dimensions
completely parallel the integration-by-parts rules of the Chern--Simons
theory in  3 dimensions \cite{CGMM}.
When we apply the functional derivative
$\displaystyle{{{\delta}\over{\delta B^a_{\rho\tau}(x)}}}$
to a surface operator $O(\Sigma';x',{x'}')$ for a surface $\Sigma'$ whose
boundary includes $C$
\footnote{We are considering here only the
small variations of $\Sigma'$, so we can disregard the holonomies
of the paths $\gamma$'s in the surface operators; the only
relevant contribution
to be considered is the one given by by the field $B=\sum B^a R^a$.}
we obtain:
\be  \displaystyle{{{\delta}\over{\delta
B^a_{\rho\tau}(x)}}}O(\Sigma';x',{x'}')
=R^a \delta^{(4)}(x-x')d^2{\sigma'}^{\rho,\tau} O(\Sigma';x',{x'}'),
\label{variation} \ee
where $d^2\sigma'$ is the surface 2-form of $\Sigma'$.

{}From eqn (\ref{intbyparts}) we conclude
that only the variations $c_x$ for which $d^2{\sigma'}\wedge d^2c_x\neq 0$
matter in the functional integral.
But these variations are exactly the ones for which
the small loop $c_x$ is such that the surface
element $d^2c_x$ is transverse to the knot $C$
(as in \cite{CGMM}).

Now we come back to the link $L_{t_l}$ with the three crossings as in Fig.\ 1.
In order to consider the more general situation we  assume that the three
strings in Fig.\ 1 belong to different components $K^1_l,\;K^2_l,\;K^3_l$
of $L_{t_l}$ that will evolve into three different
components $K^1_{l+1},\;K^2_{l+1},\;K^3_{l+1}$
of $L_{t_{l+1}}$.
By performing three such
small variations at each one of the crossings of the three strings considered
in Fig.\ 1, we can switch each under-crossing into a over-crossing.
Now as in \cite{CGMM} we compare the expectation value of the original
configuration with the configuration with the three crossings switched.
Such comparison, via the aid of eqns (\ref{intbyparts},\ref{variation})
and the Fierz identity \cite{CGMM},
leads to the conclusion \footnote{Details will be discussed elsewhere.}
that when we consider the
action of the surface operators $O(\Sigma{^{j,j}_l};x^j_l,x^j_{l+1}),$
$\;j=1,2,3,$
on the quantum holonomies of
the knots $K_l^{j}$, its infinitesimal variation $\delta O$
can be expressed as follows:
\be (\delta O)|\cdots W_l\cdots |0\rangle = \Rscr |\cdots W_l\cdots |0\rangle.
\label{infvariation} \ee
Here $W_l$ is a short notation for the operator
\[   W_l\equiv \Hol(A,K_l^{1})_{x^1_l}O(\Sigma{^{1,1}_l};x^1_l,x^1_{l+1})
\Hol(A,K_{l+1}^{1})_{x^1_{l+1}} \]
\[   \otimes \Hol(A,K_l^{2})_{x^2_l}O(\Sigma{^{2,2}_l};x^2_{l},x^2_{l+1})
\Hol(A,K_{l+1}^{2})_{x^2_{l+1}} \]
\[    \otimes
\Hol(A,K_l^{3})_{x^3_l}O(\Sigma{^{3,3}_l};x^3_{l},x^3_{l+1})
\Hol(A,K_{l+1}^{3})_{x^3_{l+1}};$$
the dots in (\ref{infvariation}) involve the operators
$W^{\rm in}(\Sigma, t_l)$ and  $W^{\rm out}(\Sigma, t_l)$
(see (\ref{W})), the vector space associated to the string $j$
is given by a tensor product $V^j_a\otimes V^j_b$, and finally
$\Rscr$ is a suitable representation of the following
operator:
$$ (1 +(1/\lambda)\sum_a R^a \otimes R^a+ \cdots)_{1a,2a}
(1 +(1/\lambda)\sum_a R^a \otimes R^a+ \cdots)_{1b,3a}$$
$$(1 +(1/\lambda)\sum_a R^a \otimes R^a+ \cdots)_{2b,3b}.$$
Now we recall that $1 + (1/\lambda)\sum_a R^a\otimes R^a$
is the infinitesimal
approximation of a quantum Yang-Baxter matrix $R$. Hence the calculations
of this section suggest that, at the finite level, the solution
of the Zamolodchikov tetrahedron equation which may be more relevant
to topological quantum field theories, is
the one considered
in \cite{CS4,Lawrence},
 namely:
\[ S_{123}=R_{1a,2a} R_{1b,3a} R_{2b,3b},\]
where $R$ is a solution of the quantum Yang-Baxter equation.

\section*{Acknowledgments}

S.\ Carter and M.\ Saito have kindly explained to us many aspects
of their work on
$2-$knots, both in person and via e-mail.
We thank them very much.
We also thank
A.\ Ashtekar, J.\ Baez, R.\ Capovilla, L.\ Kauffman, and J.\ Stasheff
very much for discussions and comments.
Both of us would also like
to thank J.\ Baez for having organized a very interesting
conference, for having invited us to participate and for helpful
comments on the manuscript.  P.\ C.-R.\ would like
to thank S.\ Carter and R.\ Capovilla for invitations to the University
of Mobile (Alabama) and Cinvestav (Mexico), respectively.

\end{document}